# 125 Gbps Pre-Compensated Nonlinear Frequency-Division Multiplexed Transmission

Son T. Le*, Vahid Aref and Henning Buelow

Nokia-Bell-Labs, Stuttgart, Germany, *son.thai_le@nokia-bell-labs.com

**Abstract** *Record-high data rate of 125 Gb/s and SE over 2 bits/s/Hz in burst-mode single-polarization NFDM transmissions were achieved over 976 km of SSMF with EDFA-only amplification by transmitting and processing 222 32 QAM-modulated nonlinear subcarriers simultaneously.*

## Introduction

Nonlinear frequency division multiplexed (NFDM) transmission has attracted a great attention recently due to the linear propagation rules of nonlinear (discrete and continuous) spectral modes (nonlinear subcarriers) over the lossless fiber channel [1-10]. There is a high expectation that the conventional Kerr-nonlinearity limit can be overcome by modulating and multiplexing signals in the nonlinear spectral domain.

During the last few years, several important experimental efforts in NFDM transmissions have been reported. In [4], an NFDM system with 7 QPSK-modulated discrete spectral modes (also called eigenvalues) providing a gross data rate of 7 Gb/s was demonstrated over 1440 km. The first NFDM system modulating both discrete and continuous modes at 26.3 Gb/s over 1464 km has been reported in [8]. Very recently, a gross data rate of 32 Gb/s has been achieved by transmitting 64 32 QAM-modulated nonlinear continuous subcarriers over 1464 km [9]. Besides, the NFDM system demonstrated in [9] also shows 1.3 dB performance gain over its linear counterpart, indicating a great potential of NFDM transmission. However, due to many yet unsolved challenges in both design and implementation, NFDM systems up to date can operate only at relatively low data rates with spectral efficiencies (SE) below 0.8 bits/s/Hz.

In this paper, we show that the data rate and the SE of NFDM transmissions can be dramatically increased by increasing the number of modulated nonlinear modes and employing a novel transmitter-side pre-compensation technique. By transmitting and processing simultaneously 222 pre-compensated 32QAM-modulated nonlinear continuous subcarriers on a single polarization, we have achieved a record-high data rate of 150 Gb/s over 976 km of SSMF, offering a net data rate of 125 Gb/s and an SE over 2 bits/s/Hz after soft-decision forward error correction (SD-FEC) decoder. In addition, a superior performance of the considered NFDM system over its linear OFDM counterpart was achieved over a wide range of nonlinear subcarrier number, from 64 to 222.

## Experimental setup

The schematic of the experimental setup, together with the Tx, Rx DSPs is shown in Fig. 1. In this experiment, the transmitted signal was designed by multiplexing $N$ overlapping sinc-shape subcarriers in the nonlinear continuous spectrum domain similarly to the way that the conventional OFDM signal spectrum is synthesized. For each data block (after serial-to-parallel conversion) and predefined launch power value, the pre-compensated multiplexed nonlinear spectrum in the normalized form was defined as:

$$q_c(\xi) = A e^{-i\xi^2 L} \cdot \sum_{k=-N/2}^{N/2} c_{m,k} \frac{\sin(\xi T_0 + k\pi)}{\xi T_0 + k\pi} e^{-2im\xi T_1},$$

where $A$ is the power control parameter, $L$ is the transmission distance, $\xi$ is the nonlinear frequency, and $\exp(-i\xi^2 L)$ is the pre-compensation of half of the transmission-induced phase shift in the nonlinear spectral domain, which is equivalent to performing 50% Tx-side digital back-propagation. The summation

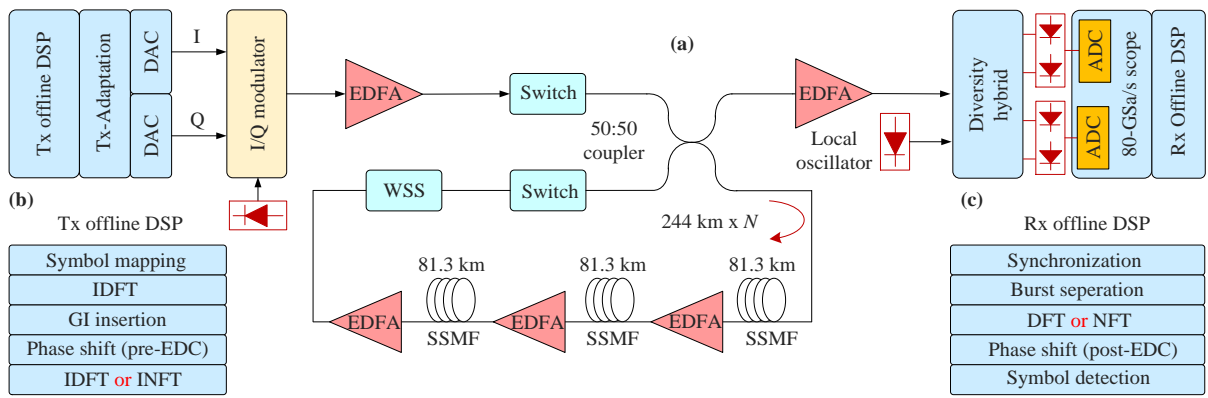

Fig. 1 (a): Experimental setup of pre-compensated NFDM and OFDM transmissions with 32 QAM format over SSMF with EDFA-only amplification; (b - c): Transmitter (Tx) and receiver (Rx) DSPs; DFT (IDFT) – discrete (inverse) discrete Fourier transforms; NFT (INFT) – nonlinear (inverse) nonlinear Fourier transforms; EDC – electrical dispersion compensation; GI – guard interval.

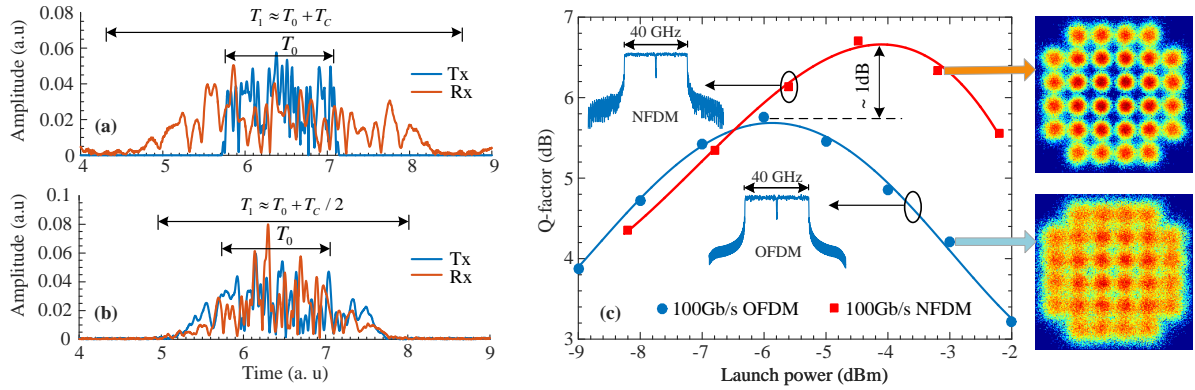

Fig. 2 (a-b): Illustration of the signal broadening during propagation in NFDM transmissions without (a) and with (b) pre-compensation technique where $T_c$ is the channel memory due to the interplay of dispersion and nonlinearity during fiber propagation, for quasi-linear transmission we have $T_c \approx 2\pi|\beta_2|BL$, where $\beta_2$ is the chromatic dispersion coefficient, $B$ is the signal bandwidth; (c) – Performance comparison of 100 Gb/s pre-compensated NFDM and OFDM systems, insets show the linear spectra of NFDM and OFDM signals.

describes the synthesized NFDM nonlinear spectrum in which $m$ is the burst index, $c_{m,k}$ is the symbol sequence drawn from 32 QAM constellation, $T_0$ is the initial burst duration before pre-compensation (blue curve, Fig. 2a), $T_1$ is the total burst duration, taking into account the signal broadening due to the pre-compensation task.

It is clear that the considered NFDM system is the nonlinear counterpart of the conventional 50% pre-electrical dispersion compensated (pre-EDC) OFDM system with the linear spectrum defined as $F(f) = q_c(\pi f)$. As a result, the generation of such NFDM signals can be effectively implemented using the inverse discrete Fourier transform (IDFT) and inverse nonlinear Fourier transform (INFT) together with the conventional pre-EDC technique as indicated in the Fig. 1b. The parameters for the considered NFDM and OFDM systems are the same and summarized in Tab. 1. Without loss of generality, we vary the number of nonlinear subcarriers from 64 to 222 and the total signal bandwidth from 32 GHz to 60 GHz while keeping the ratio $\text{GI}/T_0$ constant ($\text{GI}/T_0 = 1$, except for the case of $N = 64$). In each case, the parameters, i.e. $T_0$ and $B$, are optimized constraint to the available hardware and the NFT algorithms.

After pre-compensation, burst-mode NFDM and OFDM time-domain signals were generated using either the INFT or the conventional IDFT. To pre-compensate for both linear and nonlinear responses of the transmitter, including drivers, DAC (88 GS/s with ~ 16 GHz of bandwidth and ~5.5 bits of effective resolution) and optical modulator, an iterative adaptation routine was applied in the B2B configuration as described in [9]. This Tx adaptation routine allows highly precise optical waveforms to be generated which is critical for minimizing the implementation penalty in NFDM transmission.

The re-circulating loop consists of 3×81.3 km spans of SSMF and EDFA-only amplification. Both the transmitter laser and local oscillator were fibre lasers with ~1 kHz of linewidth. At the receiver, after coherent detection, digital sampling at 80 GS/s, timing synchronization and frequency offset compensation; the received signal was separated into bursts for further processing. Next, for OFDM and NFDM systems, DFT and NFT were performed to recover the linear and nonlinear spectrum. After that, a phase-shift operation was performed to remove the impact of dispersion in OFDM systems or the interplay of dispersion and nonlinearity in NFDM systems. Finally, channel equalization, phase noise estimation and symbol detection were carried out according to [9]. The system performance was then analyzed by the pre-FEC BER as well as the post-FEC BER estimated by processing ~ $10^6$ bits, and also by the Q-factor derived directly from the pre-FEC BER.

**Experimental results and discussion**

The performance comparison of 100 Gb/s pre-compensated NFDM transmission with its linear OFDM counterpart over 976 km is depicted in Fig. 2c. It is noted that the NFDM system offers around 1 dB performance advantage over the conventional OFDM system, indicating the effectiveness of the nonlinear compensation scheme in the nonlinear spectral domain. The pre-FEC BERs of NFDM and OFDM systems at the optimum launch power values are compared

| $T_0$ | GI | $N$ | Gross data rate | Bandwidth | SE of NFDM Systems |
|---|---|---|---|---|---|
| 2 ns | 4 ns | 64 | 53 Gb/s | 32 GHz | 1.56 bits/s/Hz |
| 3.3 ns | 3.3 ns | 132 | 100 Gb/s | 40 GHz | 2.30 bits/s/Hz |
| 3.5 ns | 3.5 ns | 154 | 110 Gb/s | 44 GHz | 2.24 bits/s/Hz |
| 3.5 ns | 3.5 ns | 176 | 125 Gb/s | 50 GHz | 2.17 bits/s/Hz |
| 3.6 ns | 3.6 ns | 198 | 137 Gb/s | 55 GHz | 2.12 bits/s/Hz |
| 3.7 ns | 3.7 ns | 222 | 150 Gb/s | 60 GHz | 2.08 bits/s/Hz |

Tab. 1 Parameters of investigated NFDM and OFDM systems. The achieved SE of NFDM systems over 976km are also given.

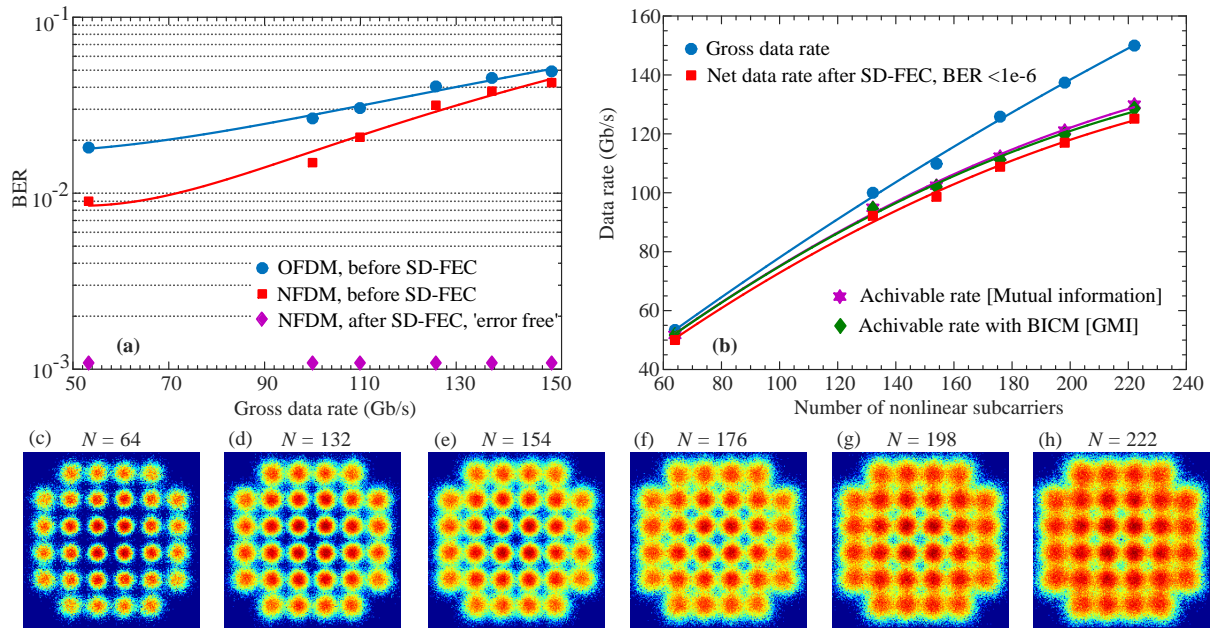

Fig. 3 (a): Pre-FEC BER vs. the gross data rate for the considered OFDM and NFDM systems after propagation over 976 km. After SD-FEC decoder with 20% overhead all considered NFDM systems are "error free" (BER<$10^{-6}$). (b) – Gross and net data rates of NFDM systems with variable overhead SD-FEC vs. the number of nonlinear subcarriers. (c-h) –Received constellations in the NFDM systems.

in Fig. 3a, showing that the performance benefit of NFDM over OFDM, which was already observed at lower data rate [9], is maintained when the gross data rate is increased to 150 Gb/s (with 222 subcarriers). We also applied soft-decision FECs to reliably recover the transmitted bits. We used spatially coupled LDPC codes (SC-LDPC) with an efficient windowed decoder [10]. FEC decoding was emulated by offline processing, without the necessity of actually implementing the FEC encoder while preparing waveforms at the Tx side for measurement. For all considered NFDM systems, error-free transmission (with ~ $10^6$ bits processed per measurement) over 976 km was achieved after SD-FEC decoder with 20% overhead. This result indicates that a record-high net data rate of 125 Gb/s has been achieved with 222 32 QAM-modulated nonlinear subcarriers.

We also noted that lower BER was observed with the reducing of the modulated nonlinear subcarriers number. This is attributed to the significant increase of implementation penalty at high bandwidth signal due to the bandwidth limitation of the transceivers. For each of considered NFDM systems, we optimized the code among a class of check-regular SC-LDPC codes to maximize the achievable rates with the target post-FEC BER smaller than $10^{-6}$. As shown in Fig. 3b, we were able to decode reliably up to the net rates very close to the generalized mutual information (GMI) estimations. The achieved SEs are shown in the Tab.1, indicating that a record-high SE of 2.3 bits/s/Hz was also achieved. In term of total data rate, Fig. 3b exhibits that increasing the number of nonlinear subcarriers is an effective way to increase the data rate of NFDM transmissions.

## Conclusion

By employing a simple Tx pre-compensation technique together with variable overhead SC-LDPC codes and by increasing the number of modulated nonlinear subcarriers up to 222, we have achieved record-high net data rate of 125 Gb/s and SE of 2.3 bits/s/Hz in NFDM transmissions over 976km of SSMF.

## References

[1] M. I. Yousefi et al, "Information transmission using the nonlinear Fourier transform, Parts I–III," IEEE Trans. Inf. Theory, 60 (2014).
[2] Z. Dong et al, "Nonlinear Frequency Division Multiplexed Transmissions Based on NFT," IEEE PTL, 27 (2015).
[3] V. Aref et al, "Experimental Demonstration of Nonlinear Frequency Division Multiplexed Transmission," ECOC 15, paper Tu. 1.1.2.
[4] H. Buelow et al. "Transmission of waveforms determined by 7 eigenvalues with PSK-modulated spectral amplitudes," ECOC 2016.
[5] S. T. Le et al, "Nonlinear inverse synthesis for high spectral efficiency transmission in optical fibers," Opt. Express, 22 (2014).
[6] S. T. Le et al, "Demonstration of nonlinear inverse synthesis transmission over transoceanic distances," JLT, 34. 2459-2466 (2016)
[7] T. Gui et al, "Alternative Decoding Methods for Optical Communications based on Nonlinear Fourier Transform," JLT, 99 (2017).
[8] V. Aref et al, "Demonstration of fully nonlinear spectrum modulated system in the highly nonlinear optical transmission regime," ECOC 2016, paper Th.3.B.2.
[9] S. T. Le, H. Buelow and V. Aref, "Demonstration of 64 x 0.5Gbaud Nonlinear Frequency Division Multiplexed Transmissions with 16QAM and 32QAM Formats," OFC 2017, paper W3J.1.
[10] L. Schmalen et al. "Spatially Coupled Soft-Decision Error Correction for Future Lightwave Systems," JLT, (2015).